\documentstyle[amssymb,aps,prl,epsfig]{revtex} 
\draft     
 
\begin{document}

\title{\bf Pseudo gap in the density of states in cuprates}  
\author{Peter Prelov\v sek and Anton Ram\v sak} 
\address{J. Stefan Institute,  SI-1000 Ljubljana, Slovenia \\ 
Faculty of Mathematics and Physics, University of Ljubljana, 
SI-1000 Ljubljana, Slovenia 
}

\date{\today} 
\maketitle 
                
\begin{abstract} 
\widetext 
\smallskip 
In the framework of the $t$-$J$ model for cuprates we analyze the
development of a pseudo gap in the density of states (DOS), which at
low doping starts to emerge for temperatures $T<J$ and persists up to
the optimum doping.  The analysis is based on numerical results for
spectral functions obtained with the finite-temperature Lanczos method
for finite two-dimensional clusters. We find that the pseudo gap
scales with $J$ and is robust also in the presence of nearest neighbor
repulsive interaction. Numerical results are additionally compared
with the self consistent Born approximation (SCBA) results for
hole-like (photoemission) and electron-like (inverse photoemission)
spectra at $T=0$. The analysis is suggesting that the origin of the
pseudo gap is in short-range antiferromagnetic (AFM) spin correlations
and strong asymmetry between the hole and electron spectra in the
underdoped regime.
\end{abstract}
\baselineskip = 2\baselineskip  % double space the text 
\vskip 1 cm

In this paper we present the theoretical analysis of DOS in planar
cuprates. As a prototype model we take the standard $t$-$J$ model,
which incorporates strong electron correlations leading to AFM in
undoped material and hindered motion of holes in doped system. The
emphasis of the present study is on the pseudo gap found in recent
angle-resolved photoemission (ARPES) experiments \cite{exp} and also
in some exact diagonalization studies \cite{jpspec,rev,sigmac}.  We
add to the model also nearest neighbor repulsion $V$ term,
\[
H = -t   \sum_{\langle ij\rangle s} ( c^\dagger_{js}c_{is}  +  
{\rm{H.c.}}) +   \sum_{\langle ij\rangle}
[ J {\bf S}_{i} \cdot  {\bf
S}_{j}  +  (V -               {J\over 4}) n_{i} n_{j}].
\]
Here $i,j$ refer to planar sites on a square lattice and
$c_{is},c^\dagger_{is}$ represent projected fermion
operators forbidding double occupation of sites.

We study here the planar DOS, defined as ${{\cal N}}(\omega) = 2/ N
\sum_{{\bf k}} { A}({\bf k},\omega-\mu)$, where $A({\bf k},\omega)$ is
the electron spectral function \cite{jpspec}, and $\mu$ denotes the
chemical potential. First we calculate the DOS with the
finite-temperature Lanczos method \cite{rev} for clusters of $N=18,
20$ sites doped with one hole, $N_h=1$. Note that ${{\cal
N}^-}(\omega)$ corresponds to adding a hole into the system and thus
to the photoemission experiments, while ${{\cal N}^+}(\omega)$
represents the inverse photoemission (IPES) spectra.
     
In Fig.~1 we present  ${{\cal N}}(\omega)$ for $J/t=0.3, 0.6$
\cite{jpspec,sigmac} on a $N=18$ sites cluster. We note that the
pseudo gap scales approximately as $2J$. The analysis at elevated
temperatures shows that the gap slowly fills up and disappears at $T
\sim J$. The gap remains robust also in the presence of the repulsive $V$
term, which on the other hand suppresses binding of hole pairs. Such
an analysis thus suggests that the origin of the pseudo gap is in
short-range AFM spin correlations rather than in the binding tendency
of doped holes.
%\renewcommand{\topfraction}{1.}        % najvec 70% prostora za slike
%na vrhu strani   
%\setcounter{bottomnumber}{3}           % najvec 1 slika na dnu strani
%\renewcommand{\bottomfraction}{1}      % najvec 30% prostora za slike 
%na dnu strani
%\setcounter{totalnumber}{3}            % najvec 3 slike
%\renewcommand{\textfraction}{0.}       % najmanj 20% besedila
%\renewcommand{\floatpagefraction}{1.}  % pol strani je lahko prazne

In Fig.~2(a) are shown spectra $\,\,{\cal N}(\omega)\,\,$ obtained
on a $\,N\,=\,20\,$ sites cluster.   We $\,\,$ compare these\hfill
spectra with the DOS within the self-consistent Born approximation
\cite{sch88}, obtained in the following manner.  We assume that ${\cal
N}^-(\omega)$ can be approximated with the SCBA {\it hole} Green's
function for adding a hole to and antiferromagnetic reference system
\cite{sch88}, Fig.~2(b).  ${\cal N}^+(\omega)$ can be in SCBA
correctly calculated and is presented in Fig.~2(c). The peaks in
${\cal N}^+(\omega)$ can well be explained with magnon structure of
single hole ground state
\cite{ramsak93}. As seen in Fig.~2(a) is the total DOS obtained
with the SCBA (dashed lines) a reasonable approximation of numerical
results.

We conclude stressing that the origin of the pseudo gap found in
cuprates seems to be in the short range spin correlations of the
reference AFM system, as well as in the strong asymmetry between the
hole-like and electron-like spectra in underdoped systems. Namely,
${\cal N}^+(\omega)$ should scale linearly with doping $c_h$ but not
changing substantially the width and form, while ${\cal N}^-(\omega)$
away from chemical potential is less sensitive to $c_h$.  Since $\mu$
lies in the pseudo gap, it is plausible that the pseudo gap observable
in ARPES should also fill up with $c_h$, as found in experiments
\cite{exp}.

\vskip -1 cm
\begin{figure}[hhh]
\center{\epsfig{file=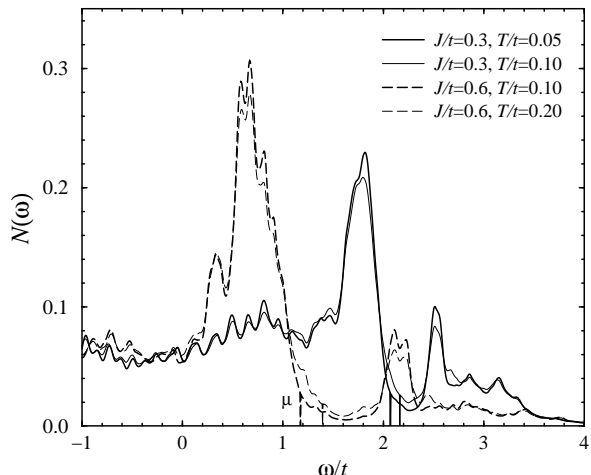,height=90mm,angle=-90}}

\caption{${\cal N}(\omega)$ for one hole on $N=18$ sites, presented
for different $J/t$ and $T/t$, at $V=0$.
Broadening of peaks is taken $\epsilon/t=0.04$.}
\end{figure}
\newpage
\begin{figure}[Htb]
\vskip -2.3 cm \hskip 0 cm          
\center{\epsfig{file=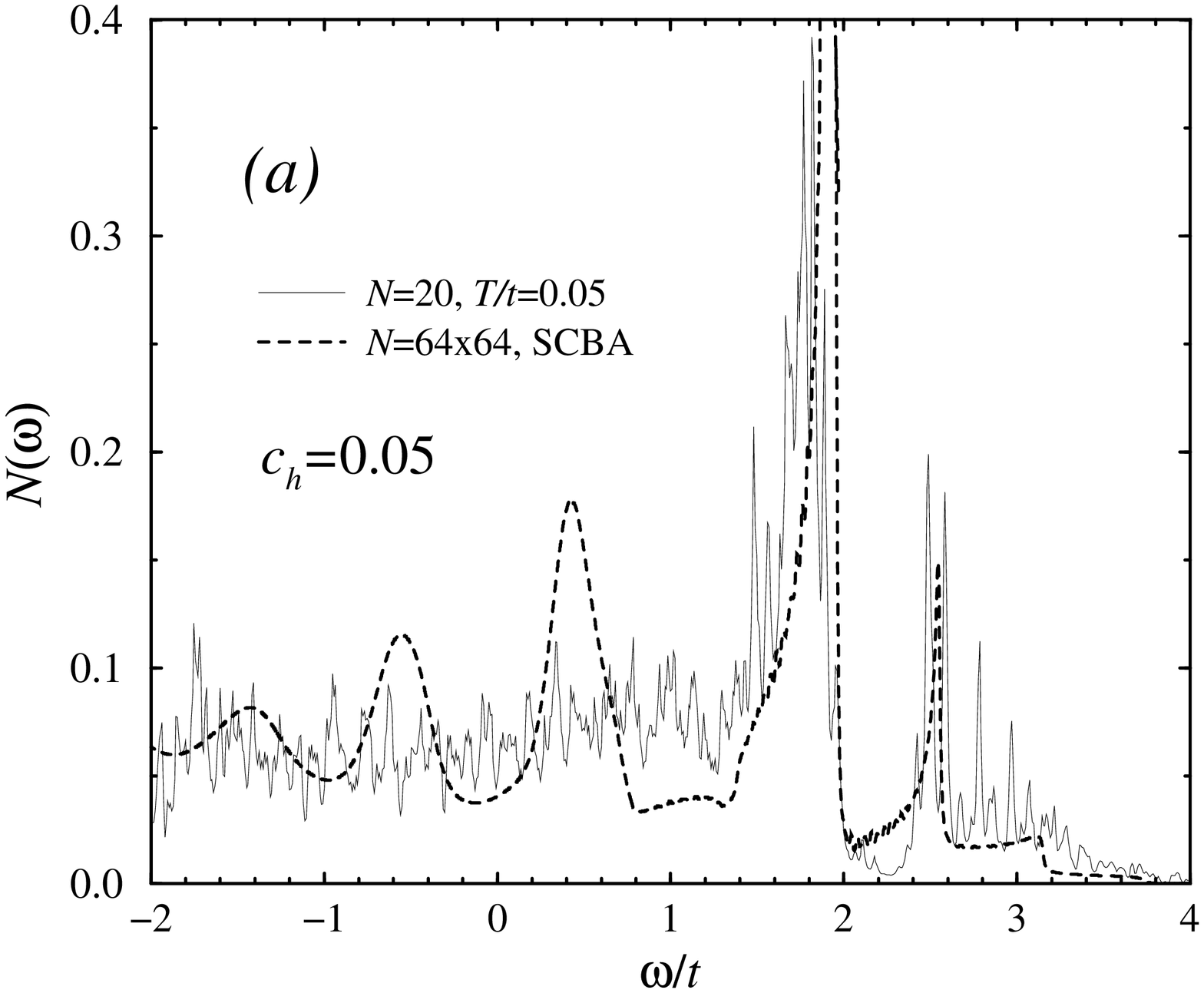,height=110mm,angle=-90}}
\vskip -3.6 cm \hskip 0 cm    
\center{\epsfig{file=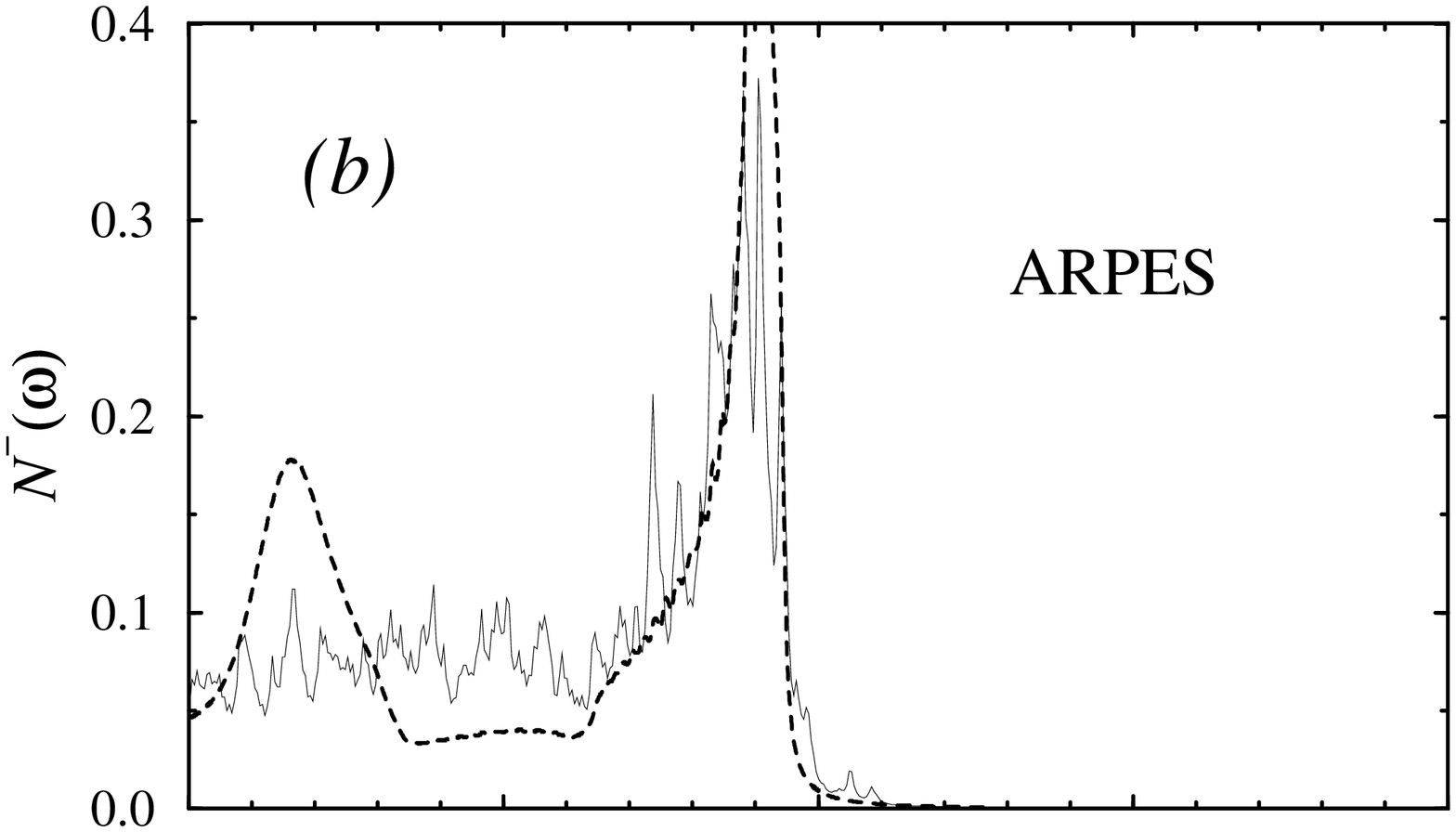,height=110mm,angle=-90}}
\vskip -5.65 cm \hskip 0 cm
\center{\epsfig{file=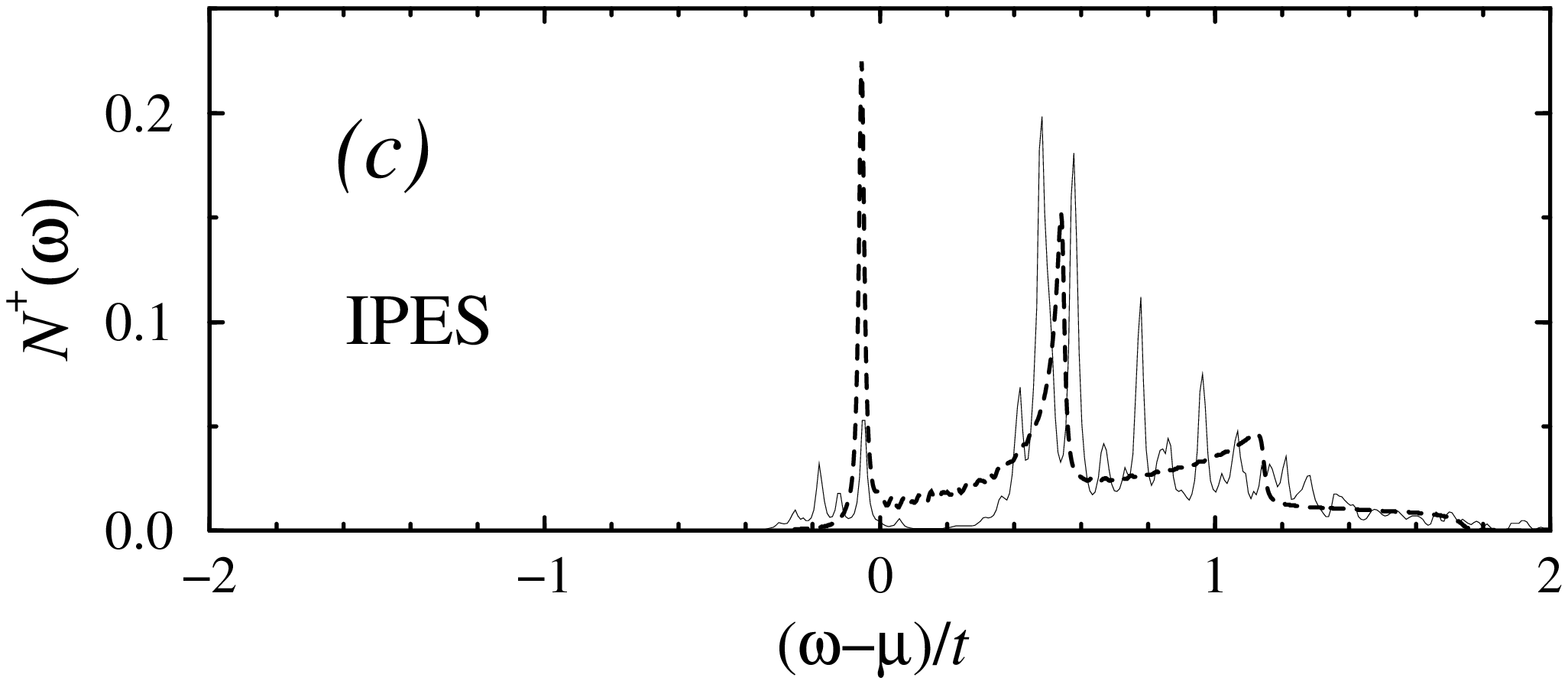,height=110mm,angle=-90}}

\caption{(a) ${\cal N}(\omega)$ for $N_h=1$ on $N=20$ sites, with
$J/t=0.3$, $V=0$, $T/t=0.05$ (full line). Dashed heavy lines represent
the SCBA result on large lattice obtained as a sum of ${\cal
N}^-(\omega)$ and ${\cal N}^+(\omega)$, presented below. (b) Hole-like
${\cal N}^-(\omega)$ spectra. The SCBA result is obtained on a $N=64
\times 64$ cluster and for {\it undoped} reference system. Note the
'string states' resonances, absent in the finite doping Green's
function. (c) ${\cal N}^+(\omega)$ corresponding to IPES. Reference
hole concentration is $c_h=1/N$. The SCBA result is normalized to
$c_h=1/20$. Broadening of peaks is taken $\epsilon/t=0.01$.}
\end{figure}

\end{document}